\documentclass[12pt]{article}
\usepackage[english]{babel}
\usepackage{amsmath,amssymb}

\begin{document}

\newtheorem{theorem}{\bf Theorem}[section]
\newtheorem{lemma}[theorem]{\bf Lemma}
\newtheorem{corollary}[theorem]{\bf Corollary}
\newtheorem{assumption}[theorem]{\bf Assumption}

\def\proof{\noindent\bf Proof. \rm}

\def\Im{\mathop{\rm Im}\nolimits}
\def\rot{\mathop{\rm rot}\nolimits}
\def\div{\mathop{\rm div}\nolimits}
\def\dom{\mathop{\rm Dom}\nolimits}
\def\supp{\mathop{\rm supp}\nolimits}

\title{Homogenisation for a statinory Maxwell system}

\author{Pozharskii Alexey A.\footnote{St. Petersburg State University, Physical Faculty,
Department of Mathematical Physics, St. Petersburg, Russia; e-mail: pozharsky@math.nw.ru.}
\thanks{The work was partially supported by the grants RFBR 07-01-92169-\_a and 08-01-00209-a.}}

\maketitle

\begin{abstract}
We study homogenization problem for the stationary Maxwell system.
It is supposed that the magnetic permeability and the dielectric permittivity
locally close to fast-oscillating (with respect to some small parameter)
periodic functions which can change the form on rather big distances.
An asymptotic behavior to solutions of the Maxwell system outside of its spectrum is obtained.
We also describe asymptotic behavior of resolvent with control of the remainder
in terms of some appropriate operator norms.
\end{abstract}


\section{Introduction}
\setcounter{equation}{0}

The work is devoted to study the homogenization problem for the Maxwell operator.
Roughly speaking it is supposed that the magnetic permeability and the dielectric permittivity locally close to fast-oscillating periodic functions which can change the form on rather big distances.
Problems of this kind for operators with purely periodic coefficients were intensively studied by many authors
\cite{BaPa,Bi4,Bi5,Ka1,Ol,Sa,Zh2}
The Maxwell operator with purely periodic coefficients was in details studied in a series of works of M.Sh.Birman and T.A.Suslina, last results can be found in work \cite{Su1}.
However, homogenization problems concerned to operators whose coefficients aren't purely periodic are investigated less in details \cite{Bor1,Bu1}.
Moreover, it seems like, the Maxwell operator with coefficients locally close to fast-oscillating periodic functions remains almost not studied.

Let us denote by $L_2(\mathbb{R}^{3};\mathbb{C}^{3})$ the $L_2$ class of $\mathbb{C}^{3}$-valued functions in $\mathbb{R}^{3}$ and
by $H^p(\mathbb{R}^{3};\mathbb{C}^{3})$ the corresponding Sobolev classes of order $p$, where $p\in\mathbb{N}$.
Let us put
\begin{equation*}
J = \{ u : u \in L_2(\mathbb{R}^{3};\mathbb{C}^{3}),\ \div(u) = 0 \}.
\end{equation*}
Here equality $\div(u) = 0$ is understood in sense of distribution theory
\begin{equation*}
\div(u) = 0 \Longleftrightarrow
\int\limits_{\mathbb{R}^3} (u, \nabla w) \, d x = 0
\quad \forall \quad w \in H^1(\mathbb{R}^{3};\mathbb{C}^{3}),
\end{equation*}
where $(\cdot, \cdot)$ is the standard scalar product in $\mathbb{C}^{3}$.

Let us consider the model described by the Maxwell operator
\begin{equation*}
\mathfrak{M}
\begin{pmatrix}
u \\ v
\end{pmatrix} =
\begin{pmatrix}
0 & i \rot(\mu^{-1} \cdot) \\
-i \rot(\alpha^{-1} \cdot) & 0
\end{pmatrix}
\begin{pmatrix}
u \\ v
\end{pmatrix},
\end{equation*}
where $\alpha$ and $\mu$ are some matrix-valued functions which will be described in detail lately.
The Maxwell operator $\mathfrak{M}$ acts in the space $J\oplus J$ on the domain
\begin{equation*}
\dom{\mathfrak{M}} = \{u, v :
u, v \in H^1(\mathbb{R}^3, \mathbb{C}^3), \quad \div(u) = 0, \quad \div(v) = 0 \}.
\end{equation*}

Now let us describe $\alpha$ and $\mu$.
Let $\Gamma$ be a lattice in $\mathbb{R}^3$ and $\Omega$ be the elementary cell of the lattice $\Gamma$.
Let coefficients $\alpha$ and $\mu$ have the form
\begin{equation*}
\alpha = \alpha\left(x, \frac{x}{\varepsilon}\right),
\mu = \mu\left(x, \frac{x}{\varepsilon}\right),
\end{equation*}
where $\varepsilon$ is a small positive parameter.
Suppose that the matrix-valued functions $\alpha$ and $\mu$ satisfy the following assumption.

\begin{assumption}\label{assAlMu}\

\noindent
1) $\alpha$ and $\mu$ are real and positive-definite $(3\times 3)$ matrixes;

\noindent
2) $\alpha, \mu \in C^\infty(\mathbb{R}^6, M_3)$;

\noindent
3) $\alpha(x, y)$ and $\mu(x, y)$ are $\Gamma$-periodic with respect to $y$;

\noindent
4) $\alpha(x, y) \equiv I$ and $\mu(x, y) \equiv I$ for any $x \not\in B_R = \{|x| < R\}$ and $y\in\mathbb{R}^3$,
where $I$ is the identity matrix and $R$ is a positive constant.
\end{assumption}

Denote by $\mathfrak{M}(\varepsilon)$ the Maxwell operator with coefficients $\alpha$ and $\mu$ satisfying assumption \ref{assAlMu}.
The Maxwell operator $\mathfrak{M}(\varepsilon)$ is closed in $J\oplus J$ with the standard scalar product and
selfadjoint in $J\oplus J$ with scalar product of the form
\begin{equation*}
\left<\begin{pmatrix}
u_1 \\ v_1
\end{pmatrix},
\begin{pmatrix}
u_2 \\ v_2
\end{pmatrix}
\right>
=
\int\limits_{\mathbb{R}^3} (\alpha^{-1} u_1, u_2) + (\mu^{-1} v_1, v_2) \, d x.
\end{equation*}

We study an asymptotic behaviour as $\varepsilon\rightarrow 0$ of the resolvent $(\mathfrak{M}(\varepsilon) - E )^{-1}$ for $\Im{E} > 0$.
We also investigate asymptotic behaviour of solutions to the Maxwell system
\begin{equation}\label{MainEqXEps}
(\mathfrak{M}(\varepsilon) - E )
\begin{pmatrix}
U(x, \varepsilon) \\ V(x, \varepsilon)
\end{pmatrix} =
\begin{pmatrix}
f^u(x) \\ f^v(x)
\end{pmatrix},\quad
\begin{pmatrix}
U(x, \varepsilon) \\ V(x, \varepsilon)
\end{pmatrix}
\in \dom{\mathfrak{M}(\varepsilon)},
\end{equation}
where $\Im{E} > 0$ and
\begin{equation}\label{AssFuv}
f^u, f^v \in C^\infty(\mathbb{R}^3, \mathbb{C}^3) \cap L_2(\mathbb{R}^3, \mathbb{C}^3),\
\div(f^u) = 0,\ \div(f^v) = 0.
\end{equation}

The main results of the work are contained in theorem \ref{ThMain} and corollary \ref{ClrMain}.

\section{Formal asymptotic solutions}
\setcounter{equation}{0}

This section is devoted to the constructing of formal asymptotic solutions, as $\varepsilon\rightarrow 0$,
of the system (\ref{MainEqXEps}).
To separate the slow and fast dependencies on the argument we seek the solution of system  (\ref{MainEqXEps}) in the form
\begin{equation}
\label{MainRep}
\begin{pmatrix}
U(x, \varepsilon) \\ V(x, \varepsilon)
\end{pmatrix} =
\begin{pmatrix}
u(x, x/\varepsilon, \varepsilon) \\ v(x, x/\varepsilon, \varepsilon)
\end{pmatrix}.
\end{equation}
It is easy to see that the following lemma holds.

\begin{lemma}
Let functions $u$ and $v$ satisfy equations
\begin{equation}\label{MainReqEqEps}
(\varepsilon^{-1} \mathfrak{M}_y +  \mathfrak{M}_x - E )
\begin{pmatrix}
u(x, y, \varepsilon) \\ v(x, y, \varepsilon)
\end{pmatrix} =
\begin{pmatrix}
f^u(x) \\ f^v(x)
\end{pmatrix},
\end{equation}
\begin{equation}\label{MainReqBCEps}
\varepsilon^{-1} \div_y u(x, y, \varepsilon) = \div_x u(x, y, \varepsilon),
\quad
\varepsilon^{-1} \div_y v(x, y, \varepsilon) = \div_x v(x, y, \varepsilon),
\end{equation}
where
\begin{equation*}
\mathfrak{M}_x
\begin{pmatrix}
u(x, y) \\ v(x, y)
\end{pmatrix} =
\begin{pmatrix}
i \rot_x(\mu^{-1}(x, y) v(x, y)) \\
-i \rot_x(\alpha^{-1}(x, y) v(x, y))
\end{pmatrix},
\end{equation*}
\begin{equation*}
\mathfrak{M}_y
\begin{pmatrix}
u(x, y) \\ v(x, y)
\end{pmatrix} =
\begin{pmatrix}
i \rot_y(\mu^{-1}(x, y) v(x, y)) \\
-i \rot_y(\alpha^{-1}(x, y) v(x, y))
\end{pmatrix}
\end{equation*}
and the operators $\rot_x$, $\div_x$ and $\rot_y$, $\div_y$ act upon a variables $x$ and $y$ respectively.
Then a function
\begin{equation*}
\begin{pmatrix}
U(x, \varepsilon) \\ V(x, \varepsilon)
\end{pmatrix}
=
\begin{pmatrix}
u\left(x, \frac{x}{\varepsilon}, \varepsilon\right) \\
v\left(x, \frac{x}{\varepsilon}, \varepsilon\right)
\end{pmatrix}
\end{equation*}
satisfies equation {\rm (\ref{MainEqXEps})} and conditions
$\div(U(x, \varepsilon)) = 0$, $\div(V(x, \varepsilon)) = 0$.
\end{lemma}

\proof
It can be easily proved by using direct calculations.~$\square$

Thus, the constructing of formal solutions of system (\ref{MainEqXEps}) is reduced to the constructing of formal solutions of equations (\ref{MainReqEqEps}), (\ref{MainReqBCEps})

The formal solutions of equations (\ref{MainReqEqEps}), (\ref{MainReqBCEps}) are seeked in the form
\begin{equation}
\label{ReprAsympUV}
\begin{pmatrix}
u(x, y, \varepsilon) \\ v(x, y, \varepsilon)
\end{pmatrix} =
\sum_{n\geqslant 0} \varepsilon^n
\begin{pmatrix}
u_n(x, y) \\ v_n(x, y)
\end{pmatrix},
\end{equation}
where functions $u_n(x, y)$ and $v_n(x, y)$ are $\Gamma$-periodic with respect to $y$.

\begin{lemma}
Suppose $u_n$, $v_n$ satisfy the following recurrence system of equations
\begin{equation}\label{RekEq}
\mathfrak{M}_y
\begin{pmatrix}
u_n(x, y) \\ v_n(x, y)
\end{pmatrix}
=
\begin{pmatrix}
f^u(x) \\ f^v(x)
\end{pmatrix} \delta_{n1} +
(E - \mathfrak{M}_x)
\begin{pmatrix}
u_{n-1}(x, y) \\ v_{n-1} (x, y)
\end{pmatrix},
\end{equation}
\begin{equation}\label{RekBC}
\div_y ( u_n(x, y) ) = - \div_x ( u_{n-1} (x, y) ), \quad
\div_y ( v_n(x, y) ) = - \div_x ( v_{n-1} (x, y) ),
\end{equation}
where $u_{-1} \equiv 0$ and $v_{-1} \equiv 0$.
Then {\rm (\ref{ReprAsympUV})} formally satisfies problem {\rm (\ref{MainReqEqEps})}, {\rm (\ref{MainReqBCEps})}.
\end{lemma}

\proof
Substituting expansion (\ref{ReprAsympUV}) in system (\ref{MainReqEqEps}), (\ref{MainReqBCEps}), and comparing the coefficients corresponding to the equal powers of  $\varepsilon$, we obtain recurrence system (\ref{RekEq}), (\ref{RekBC}).~$\square$

For detailed description of solutions to system (\ref{RekEq}), (\ref{RekBC}) we recall the following well known result.

\begin{lemma}\label{LmSolRot}
There exist 3 linearly independent $\Gamma$-periodic (with respect to $y$) solutions to the system
\begin{equation*}
\left\{
\begin{array}{ll}
\rot_y(\zeta(x, y)) = 0, \\
\div_y(\alpha(x, y) \zeta(x, y)) = 0.
\end{array}
\right.
\end{equation*}
\begin{equation*}
\zeta_k(x, y) = e_k + \nabla_y \varphi_k(x, y), \quad k = 1, 2, 3,
\end{equation*}
\begin{equation*}
e_1 = (1, 0, 0)^t, \quad
e_2 = (0, 1, 0)^t, \quad
e_3 = (0, 0, 1)^t,
\end{equation*}
where $\varphi_k(x, y)$ is any $\Gamma$-periodic (with respect to $y$) solutions to the equation
\begin{equation*}
\div_y(\alpha(x, y) \nabla_y \varphi_k(x, y) + \alpha(x, y) e_k) = 0.
\end{equation*}

There exist 3 linearly independent $\Gamma$-periodic (with respect to $y$) solutions to the system
\begin{equation*}
\left\{
\begin{array}{ll}
\rot_y(\xi(x, y)) = 0, \\
\div_y(\mu(x, y) \xi(x, y)) = 0.
\end{array}
\right.
\end{equation*}
\begin{equation*}
\xi_k(x, y) = e_k + \nabla_y \psi_k(x, y), \quad k = 1, 2, 3,
\end{equation*}
where $\psi_k(x, y)$ is any $\Gamma$-periodic (with respect to $y$) solutions to the equation
\begin{equation*}
\div_y(\mu(x, y) \nabla_y \psi_k(x, y) + \mu(x, y) e_k) = 0.
\end{equation*}

Moreover, $\zeta_k, \xi_k \in C^\infty(\mathbb{R}^3, \mathbb{C}^3)$ and
$\zeta_k(x, y) = e_k$, $\xi_k(x, y) = e_k$ for $x \not\in B_R$, where $k=1,2,3$.
\end{lemma}

\proof We only need to prove that $\zeta_k(x, y) = e_k$, $\xi_k(x, y) = e_k$ for $x \not\in B_R$, where $k=1,2,3$.
It is easy to see that $\alpha(x, y) = \mu(x, y) = I$ for $x \not\in B_R$.
Hence $\varphi_k$, $\psi_k$ satisfy the following equations
\begin{equation*}
\Delta_y \varphi_k(x, y) = 0, \quad
\Delta_y \psi_k(x, y) = 0, \quad x \not\in B_R, \ k=1,2,3.
\end{equation*}
Since $\varphi_k(x, y)$, $\psi_k(x, y)$ are $\Gamma$-periodic functions
(with respect to $y$),
we see that $\varphi_k(x, y)$, $\psi_k(x, y)$ are constant for $x \not\in B_R$.
This implies the necessary assertion.~$\square$

Consider recurrence system (\ref{RekEq}), (\ref{RekBC}) for $n=0$
\begin{equation}\label{ReqSyst0}
\mathfrak{M}_y
\begin{pmatrix}
u_0(x, y) \\ v_0(x, y)
\end{pmatrix}
= 0, \quad
\div_y (u_0(x, y)) = \div_y (v_0(x, y)) = 0.
\end{equation}
Lemma \ref{LmSolRot} implies that general solution of (\ref{ReqSyst0}) has the following form
\begin{equation*}
u_0(x, y) = \alpha(x, y) \sum_{k=1}^3 a_k^0(x) \zeta_k(x, y), \quad
v_0(x, y) = \mu(x, y) \sum_{k=1}^3 b_k^0(x) \xi_k(x, y),
\end{equation*}
where $a_k^0(x)$ and $b_k^0(x)$ are arbitrary smooth functions.

Recurrence system (\ref{RekEq}), (\ref{RekBC}) for $n=1$ has the from
\begin{equation}\label{ReqSyst1}
\left\{
\begin{array}{l}
\mathfrak{M}_y
\begin{pmatrix}
u_1(x, y) \\ v_1(x, y)
\end{pmatrix}
=
\begin{pmatrix}
f^u(x) \\ f^v(x)
\end{pmatrix}
+
(E - \mathfrak{M}_x)
\begin{pmatrix}
u_0(x, y) \\ v_0(x, y)
\end{pmatrix}
, \\
\div_y (u_1(x, y)) = -\div_x (u_0(x, y)), \quad
\div_y (v_1(x, y)) = -\div_x (v_0(x, y)).
\rule{0pt}{15pt}
\end{array}
\right.
\end{equation}
Let us rewrite it in the following from
\begin{equation}\label{ReqSyst1V}
\left\{
\begin{array}{l}
\rot_y (\mu^{-1} v_1)
=
-i f^u - i E u_0 - \rot_x (\mu^{-1} v_0), \\
\div_y (v_1) = -\div_x (v_0),
\rule{0pt}{15pt}
\end{array}
\right.
\end{equation}
\begin{equation}\label{ReqSyst1U}
\left\{
\begin{array}{l}
\rot_y (\alpha^{-1} u_1)
=
i f^v + i E v_0 - \rot_x (\alpha^{-1} u_0), \\
\div_y (u_1) = -\div_x (u_0).
\rule{0pt}{15pt}
\end{array}
\right.
\end{equation}

Suppose that the following conditions hold
\begin{equation}\label{SolvCondV1}
\left\{
\begin{array}{l}
\int\limits_\Omega (-i f^u - i E u_0 - \rot_x (\mu^{-1} v_0), \xi_k) \, d y = 0, \quad k = 1, 2, 3, \\
\int\limits_\Omega \div_x (v_0) \, d y = 0,
\rule{0pt}{15pt}
\end{array}
\right.
\end{equation}
then problem (\ref{ReqSyst1V}) can be solved.
It is easy to check that
\begin{equation*}
\frac{1}{|\Omega|} \int\limits_\Omega
\left(\rule{0ex}{2ex} \xi_1(x, y), \xi_2(x, y), \xi_3(x, y) \right)^t \, d y = I,
\end{equation*}
\begin{equation*}
\begin{pmatrix}
\left(\rule{0ex}{2ex} f^u(x), \xi_1(x, y)\right)_y \\
\left(\rule{0ex}{2ex} f^u(x), \xi_2(x, y)\right)_y \\
\left(\rule{0ex}{2ex} f^u(x), \xi_3(x, y)\right)_y
\end{pmatrix}
= |\Omega| f^u(x).
\end{equation*}
Similarly,
\begin{equation*}
\left(\rule{0ex}{2ex} u_0(x, y), \xi_k(x, y)\right)_y
=
|\Omega| \sum_{p=1}^3 \Lambda^u_{kp}(x) a^0_p(x),
\end{equation*}
\begin{equation*}
\begin{pmatrix}
\left(\rule{0ex}{2ex} u_0(x, y), \xi_1(x, y)\right)_y \\
\left(\rule{0ex}{2ex} u_0(x, y), \xi_2(x, y)\right)_y \\
\left(\rule{0ex}{2ex} u_0(x, y), \xi_3(x, y)\right)_y
\end{pmatrix}
= |\Omega| \Lambda^u(x) a^0(x), \quad
a^0(x) =
\begin{pmatrix}
a^0_1(x) \\
a^0_2(x) \\
a^0_3(x)
\end{pmatrix},
\end{equation*}
where
\begin{multline}\label{LaU}
\Lambda^u_{kp}(x) = \frac{1}{|\Omega|} \left(\rule{0ex}{2ex} \alpha(x, y) \zeta_p(x, y), \xi_k(x, y)\right)_y = \\
= \frac{1}{|\Omega|} \left(\rule{0ex}{2ex} \alpha(x, y) \zeta_p(x, y), \zeta_k(x, y)\right)_y.
\end{multline}
From (\ref{LaU}) it follows that $\Lambda^u(x)$ is real and positive-definite $(3\times 3)$ matrix.

Using direct calculations, we get
\begin{equation*}
\left(\rule{0ex}{2ex} \rot_x (\mu^{-1}(x, y) v_0(x, y)), \xi_k(x, y)\right)_y =
\sum_{p=1}^3 \left(\rule{0ex}{2ex} \rot_x ( b_p^0(x) \xi_p(x, y)), \xi_k(x, y)\right)_y =
\end{equation*}
\begin{multline*}
= \sum_{p=1}^3 b_p^0(x) \left(\rule{0ex}{2ex} \rot_x (\xi_p(x, y)), \xi_k(x, y)\right)_y + \\
+ \sum_{p=1}^3 \left(\rule{0ex}{2ex} \left[\nabla_x (b_p^0(x)) \times \xi_p(x, y)\right], \xi_k(x, y)\right)_y.
\end{multline*}
It is easy to see that
\begin{equation*}
\left(\rule{0ex}{2ex} \rot_x (\xi_p(x, y), \xi_k(x, y)\right)_y = 0,
\end{equation*}
\begin{equation*}
\int\limits_\Omega \left[ \xi_p(x, y) \times \xi_k(x, y) \right] \, d y = |\Omega| \left[ e_p \times e_k \right].
\end{equation*}
Therefore,
\begin{equation*}
\left(\rule{0ex}{2ex} \rot_x v_0(x, y), \xi_k(x, y)\right)_y =
|\Omega| \sum_{p=1}^3 \left(\rule{0ex}{2ex} \left[ e_p \times e_k \right], \nabla_x (b_p^0(x)) \right),
\end{equation*}
\begin{equation*}
\begin{pmatrix}
\left(\rule{0ex}{2ex} \rot_x v_0(x, y), \xi_1(x, y)\right)_y \\
\left(\rule{0ex}{2ex} \rot_x v_0(x, y), \xi_2(x, y)\right)_y \\
\left(\rule{0ex}{2ex} \rot_x v_0(x, y), \xi_3(x, y)\right)_y
\end{pmatrix}
=
|\Omega| \rot_x b^0(x), \quad
b^0(x) =
\begin{pmatrix}
b^0_1(x) \\
b^0_2(x) \\
b^0_3(x)
\end{pmatrix}.
\end{equation*}
Thus, the first three equations from (\ref{SolvCondV1}) can be rewritten in the following form
\begin{equation*}
i \rot_x b^0(x) - E \Lambda^u(x) a^0(x) = f^u(x).
\end{equation*}
The last equation from (\ref{SolvCondV1}) can be reduced to the form
\begin{equation*}
\div_x (\Lambda^v(x) b^0(x)) = 0,
\end{equation*}
where
\begin{equation}\label{LaV}
\Lambda^v_{kp}(x) = \frac{1}{|\Omega|} \left(\rule{0ex}{2ex} \mu(x, y) \xi_p(x, y), \zeta_k(x, y)\right)_y =
\frac{1}{|\Omega|} \left(\rule{0ex}{2ex} \mu(x, y) \xi_p(x, y), \xi_k(x, y)\right)_y.
\end{equation}
From (\ref{LaV}) it follows that $\Lambda^v(x)$ is real and positive-definite $(3\times 3)$ matrix.

Similarly, one can check that solvability conditions for problem (\ref{ReqSyst1U}) can be written in the form
\begin{equation*}
- i \rot_x a^0(x) - E \Lambda^v(x) b^0(x) = f^v(x),
\end{equation*}
\begin{equation*}
\div_x (\Lambda^u(x) a^0(x)) = 0.
\end{equation*}

Let us introduce notations
\begin{equation*}
\hat{u}_0(x) = \Lambda^u(x) a^0(x), \quad \hat{v}_0(x) = \Lambda^v(x) b^0(x).
\end{equation*}
Finally, the solvability conditions for (\ref{ReqSyst1}) have the following form
\begin{equation}\label{SystAver}
\left\{
\begin{array}{ll}
(\hat{\mathfrak{M}} - E)
\begin{pmatrix}
\hat{u}_0 \\ \hat{v}_0
\end{pmatrix}
=
\begin{pmatrix}
f^u(x) \\ f^v(x)
\end{pmatrix}, \\
\rule{0ex}{3ex}
\div_x (\hat{u}_0) = 0, \quad
\div_x (\hat{v}_0) = 0,
\end{array}\right.
\quad
\hat{\mathfrak{M}} =
\begin{pmatrix}
0 & i \rot_x (\Lambda^v(x))^{-1} \\
-i \rot_x (\Lambda^u(x))^{-1} & 0
\end{pmatrix}.
\end{equation}
General solution of (\ref{ReqSyst1}) can be represented in the form
\begin{equation*}
u_1(x, y) = \tilde{u}_1(x, y) + \alpha(x, y) \sum_{k=1}^3 a_k^1(x) \zeta_k(x, y),
\end{equation*}
\begin{equation*}
v_1(x, y) = \tilde{v}_1(x, y) + \mu(x, y) \sum_{k=1}^3 b_k^1(x) \xi_k(x, y),
\end{equation*}
where $\tilde{u}_1(x, y)$, $\tilde{v}_1(x, y)$ are partial solutions of (\ref{ReqSyst1}) that are orthogonal to solutions of homogeneous equations.

Let us check that $\tilde{u}_1(x, y) \equiv 0$ and $\tilde{v}_1(x, y) \equiv 0$ for $x \not\in B_R$.
It is easy to see that
\begin{equation*}
\alpha(x, y) = \mu(x, y) = \Lambda^u(x) = \Lambda^v(x) = I, \quad
\hat{u}_0(x) = a^0(x), \quad \hat{v}_0(x) = b^0(x),
\end{equation*}
for $x \not\in B_R$.
Taking into account (\ref{SystAver}), we see that for $x \not\in B_R$ the following equations hold
\begin{equation}
\left\{
\begin{array}{l}
\rot_y (\tilde{v}_1) = 0, \\
\div_y (\tilde{v}_1) = 0,
\rule{0pt}{15pt}
\end{array}
\right.
\quad
\left\{
\begin{array}{l}
\rot_y (\tilde{u}_1) = 0, \\
\div_y (\tilde{u}_1) = 0.
\rule{0pt}{15pt}
\end{array}
\right.
\end{equation}
Recalling that $\tilde{u}_1(x, y)$, $\tilde{v}_1(x, y)$ are orthogonal to solutions of homogeneous equations, we get that
$\tilde{u}_1(x, y) \equiv 0$ and $\tilde{v}_1(x, y) \equiv 0$ for $x \not\in B_R$.

Now let us consider recurrence system (\ref{RekEq}), (\ref{RekBC}) for $n\geqslant 2$.
General solution of this system can be represented in the form
\begin{equation}\label{represUnVn}
\begin{array}{l}
u_n(x, y) = \tilde{u}_n(x, y) + \alpha(x, y) \sum\limits_{k=1}^3 a_k^n(x) \zeta_k(x, y), \\
v_n(x, y) = \tilde{v}_n(x, y) + \mu(x, y) \sum\limits_{k=1}^3 b_k^n(x) \xi_k(x, y),
\rule{0pt}{5ex}
\end{array}
\end{equation}
where $\tilde{u}_n(x, y)$, $\tilde{v}_n(x, y)$ are partial solutions of (\ref{RekEq}), (\ref{RekBC}) that are orthogonal to solutions of homogeneous system.
Coefficients $a_k^n(x)$ and $b_k^n(x)$ can be defined from the solvability conditions of system (\ref{RekEq}), (\ref{RekBC}) with $n$ substituted by $n+1$, which can be written in the form
\begin{equation}\label{SystAverN}
\left\{
\begin{array}{ll}
(\hat{\mathfrak{M}} - E)
\begin{pmatrix}
\hat{u}_n \\ \hat{v}_n
\end{pmatrix}
=
\begin{pmatrix}
F_n^u(x) \\ F_n^v(x)
\end{pmatrix}, \\
\rule{0ex}{3ex}
\div_x (\hat{u}_n) = G_n^u(x), \quad
\div_x (\hat{v}_n) = G_n^v(x),
\end{array}\right.
\end{equation}
where
\begin{equation*}
\hat{u}_n(x) = \Lambda^u(x)
\begin{pmatrix}
a^n_1(x) \\
a^n_2(x) \\
a^n_3(x)
\end{pmatrix}, \quad
\hat{v}_n(x) = \Lambda^v(x)
\begin{pmatrix}
b^n_1(x) \\
b^n_2(x) \\
b^n_3(x)
\end{pmatrix},
\end{equation*}
\begin{equation*}
F_n^u(x) =
\frac{1}{|\Omega|}
\begin{pmatrix}
\left(\rule{0ex}{2ex} f_n^u(x, y), \xi_1(x, y)\right)_y \\
\left(\rule{0ex}{2ex} f_n^u(x, y), \xi_2(x, y)\right)_y \\
\left(\rule{0ex}{2ex} f_n^u(x, y), \xi_3(x, y)\right)_y
\end{pmatrix},
\end{equation*}
\begin{equation*}
f_n^u(x, y) = E \tilde{v}_n(x, y) + i \rot_x (\mu^{-1}(x, y) \tilde{v}_n(x, y)),
\end{equation*}
\begin{equation*}
F_n^v(x) =
\frac{1}{|\Omega|}
\begin{pmatrix}
\left(\rule{0ex}{2ex} f_n^v(x, y), \zeta_1(x, y)\right)_y \\
\left(\rule{0ex}{2ex} f_n^v(x, y), \zeta_2(x, y)\right)_y \\
\left(\rule{0ex}{2ex} f_n^v(x, y), \zeta_3(x, y)\right)_y
\end{pmatrix},
\end{equation*}
\begin{equation*}
f_n^v(x, y) = E \tilde{u}_n(x, y) + i \rot_x (\alpha^{-1}(x, y) \tilde{u}_n(x, y)),
\end{equation*}

\begin{equation}
G_n^u(x) = \frac{1}{|\Omega|} \int\limits_\Omega \div_x (\tilde{u}_n(x, y)) \, d y,
\quad
G_n^v(x) = \frac{1}{|\Omega|} \int\limits_\Omega \div_x (\tilde{v}_n(x, y)) \, d y.
\end{equation}

Let us collect the obtained result in the following theorem.
\begin{theorem}\label{ThFormalRaw}
Suppose $\Im(E)>0$, $\varepsilon>0$, assumption {\rm \ref{assAlMu}} and conditions {\rm (\ref{AssFuv})} hold.
Then Maxwell system {\rm (\ref{MainEqXEps})} has asymptotic solution of the form
\begin{equation}\label{FormalRaw}
\begin{pmatrix}
U(x, \varepsilon) \\ V(x, \varepsilon)
\end{pmatrix}
=
\sum_{n\geqslant 0} \varepsilon^n
\begin{pmatrix}
u_n\left(x, \frac{x}{\varepsilon}, \varepsilon\right) \\
v_n\left(x, \frac{x}{\varepsilon}, \varepsilon\right)
\end{pmatrix},
\end{equation}
where $u_n$ and $v_n$ can be represented in form {\rm (\ref{represUnVn})}.
The corresponding coefficients can be founded from recurrence equations
{\rm (\ref{RekEq})}, {\rm (\ref{RekBC})}, and {\rm (\ref{SystAverN})}.

Moreover, $\tilde{u}_n(x, y) \equiv 0$ and $\tilde{v}_n(x, y) \equiv 0$ for $x \not\in B_R$ and $n\geqslant 0$.
\end{theorem}

The following lemma describe important estimates for the components of formal solution (\ref{FormalRaw}).
\begin{lemma}\label{LmEstForAsimp}
Suppose that the statements of theorem {\rm \ref{ThFormalRaw}} hold. Then
\begin{equation}\label{EqLm1Est1}
\max_{0 \leqslant |k| \leqslant 2} \max_{y \in \Omega}
\left\| \frac{\partial^{|k|}}{\partial y^k} u_n(\cdot, y, \varepsilon) \right\|_{H^{s-n+1}(\mathbb{R}^3, \mathbb{C}^3)}
\leqslant C \| f \|_{H^{s}(\mathbb{R}^3, \mathbb{C}^3)},
\end{equation}
\begin{equation}\label{EqLm1Est2}
\max_{0 \leqslant |k| \leqslant 2} \max_{y \in \Omega}
\left\| \frac{\partial^{|k|}}{\partial y^k} v_n(\cdot, y, \varepsilon) \right\|_{H^{s-n+1}(\mathbb{R}^3, \mathbb{C}^3)}
\leqslant C \| f \|_{H^{s}(\mathbb{R}^3, \mathbb{C}^3)},
\end{equation}
\begin{equation}\label{EqLm1Est3}
\left\| u_n(\cdot, \cdot/\varepsilon, \varepsilon) \right\|_{H^{s-n+1}(\mathbb{R}^3, \mathbb{C}^3)}
\leqslant C \varepsilon^{n-s-1} \| f \|_{H^{s}(\mathbb{R}^3, \mathbb{C}^3)},
\end{equation}
\begin{equation}\label{EqLm1Est4}
\left\| v_n(\cdot, \cdot/\varepsilon, \varepsilon) \right\|_{H^{s-n+1}(\mathbb{R}^3, \mathbb{C}^3)}
\leqslant C \varepsilon^{n-s-1} \| f \|_{H^{s}(\mathbb{R}^3, \mathbb{C}^3)},
\end{equation}
for $0 \leqslant n \leqslant s + 1$, where the constant $C$ does not depend on $f$.
\end{lemma}

\proof
The statement of the lemma for $n=0$ follows from (\ref{SystAver}).
Using recurrent system of equations (\ref{RekEq}), (\ref{RekBC}), and (\ref{SystAverN}),
the statement of the lemma for $n\geqslant 1$ can be easily proved by induction.~$\square$

\section{Asymptotic expansion of the resolvent}
\setcounter{equation}{0}

In this section we describe asymptotic behaviour of the resolvent $(\mathfrak{M}(\varepsilon) - E )^{-1}$
for a small $\varepsilon$.
To do that we consider a partial sum for formal series (\ref{FormalRaw}) of the form
\begin{equation*}
\begin{pmatrix}
U_N(x, x/\varepsilon, \varepsilon) \\ V_N(x, x/\varepsilon, \varepsilon)
\end{pmatrix}
=
\sum_{n=0}^N \varepsilon^n
\begin{pmatrix}
u_n(x, x/\varepsilon) \\ v_n(x, x/\varepsilon)
\end{pmatrix}
+
\varepsilon^N \delta(x, \varepsilon),
\end{equation*}
where $N\geqslant 1$ and $\delta = (\delta^u, \delta^v)^t$ is an auxiliary function.
Here we also suppose that $\hat{v}_N(x) \equiv \hat{u}_N(x) \equiv 0$. This implies that
$u_N(x, x/\varepsilon) \equiv v_N(x, x/\varepsilon) \equiv 0$ for $x \not\in B_R$.

It easy to see that
\begin{multline}\label{EqSevazkoj}
(\mathfrak{M}(\varepsilon) - E )
\begin{pmatrix}
U_N \\ V_N
\end{pmatrix}
=
\begin{pmatrix}
f^u(x) \\ f^v(x)
\end{pmatrix}
+ \\
+
\varepsilon^N
\left[
(\mathfrak{M}_x - E )
\begin{pmatrix}
u_N(x, y) \\ v_N(x, y)
\end{pmatrix}
\right]_{y=x/\varepsilon}
+
\varepsilon^N (\mathfrak{M}(\varepsilon) - E ) \delta(x, \varepsilon),
\end{multline}
\begin{equation}\label{EqNevU}
\div\left(U_N\right) =
\varepsilon^N \left[\div_x(u_N(x, y))\right]_{y=x/\varepsilon}
+
\varepsilon^N \div(\delta^u(x, \varepsilon)),
\end{equation}
\begin{equation}\label{EqNevV}
\div\left(V_N\right) =
\varepsilon^N \left[\div_x(v_N(x, y))\right]_{y=x/\varepsilon}
+
\varepsilon^N \div(\delta^v(x, \varepsilon)).
\end{equation}

Now we chose $\delta$ such that $(U_N, V_N)^t \in \dom{\mathfrak{M}}$.
To do that we find an appropriate solution to the following equations $\div\left(U_N\right) = 0$,
$\div\left(V_N\right) = 0$.
First equation can be rewritten in the form
\begin{equation}\label{EqDivgU}
\div(\delta^u(x, \varepsilon)) = g(x, \varepsilon) \equiv -\left[\div_x(u_N(x, y))\right]_{y=x/\varepsilon}.
\end{equation}
Let us consider the following solution of equation (\ref{EqDivgU})
\begin{equation*}
\delta^u(x, \varepsilon) = x \int\limits_0^1 g(t x, \varepsilon) t^2 \, d t.
\end{equation*}
Recalling that $\supp{g} \subset B_R$, we obtain
\begin{multline*}
\| \delta^u(\cdot, \varepsilon) \|^2_{H^1(\mathbb{R}^3, \mathbb{C}^3)} \leqslant
(R^2 + 6) \int\limits_{\mathbb{R}^3} \left( \int\limits_0^1 |g(t x, \varepsilon)| t^2 \, d t \right)^2 \, dx + \\
+ 2 R^2 \sum_{j=1}^3 \int\limits_{\mathbb{R}^3} \left( \int\limits_0^1 |g_j(t x, \varepsilon)| t^3 \, d t \right)^2 \, dx,
\end{multline*}
where $g_j(x, \varepsilon) = \partial_{x_j} g(x, \varepsilon)$.
Using simple calculations, we get
\begin{equation*}
\int\limits_{\mathbb{R}^3} \left( \int\limits_0^1 |g(t x, \varepsilon)| t^2 \, d t \right)^2 \, dx
\leqslant
\| g(\cdot, \varepsilon)\|^2_{L_2(\mathbb{R}^3,\mathbb{C}))},
\end{equation*}
\begin{equation*}
\int\limits_{\mathbb{R}^3,\mathbb{C}} \left( \int\limits_0^1 |g_j(t x, \varepsilon)| t^3 \, d t \right)^2 \, dx
\leqslant
\| g_j(\cdot, \varepsilon)\|_{L_2(\mathbb{R}^3,\mathbb{C})}^2, \quad k = 1,2,3,
\end{equation*}
and hence
\begin{multline*}
\| \delta^u(\cdot, \varepsilon)\|_{H^1(\mathbb{R}^3, \mathbb{C}^3)}^2 \leqslant
(R^2 + 6) \| g(\cdot, \varepsilon)\|_{L_2(\mathbb{R}^3,\mathbb{C})}^2 + \\
+ 2 R^2 \sum_{j=1}^3 \| g_j(\cdot, \varepsilon)\|_{L_2(\mathbb{R}^3,\mathbb{C})}^2 \leqslant
(2 R^2 + 6) \| g(\cdot, \varepsilon)\|_{H_1(\mathbb{R}^3,\mathbb{C})}^2.
\end{multline*}
Using lemma~\ref{LmEstForAsimp} and definition (\ref{EqDivgU}) of $g$ we obtain
\begin{equation}\label{EstForDeU}
\| \delta^u(\cdot, \varepsilon)\|_{H^1(\mathbb{R}^3, \mathbb{C}^3)} \leqslant
(2 R^2 + 6) \varepsilon^{-1} \| f \|_{H^{N+1}(\mathbb{R}^3, \mathbb{C}^3)}.
\end{equation}

In the same way one can define $\delta^v$ such that $\div\left(V_N\right) = 0$ and
\begin{equation}\label{EstForDeV}
\| \delta^v(\cdot, \varepsilon)\|_{H^1(\mathbb{R}^3, \mathbb{C}^3)} \leqslant
(2 R^2 + 6) \varepsilon^{-1} \| f \|_{H^{N+1}(\mathbb{R}^3, \mathbb{C}^3)}.
\end{equation}

Since $(u_N(x, y), v_N(x, y))^t \in \dom{\mathfrak{M}}$, we can apply the resolvent $(\mathfrak{M}(\varepsilon) - E )^{-1}$ to the both sides of (\ref{EqSevazkoj}) for $\Im(E) > 0$

\begin{equation*}
(\mathfrak{M}(\varepsilon) - E )^{-1}
\begin{pmatrix}
f^u(x) \\ f^v(x)
\end{pmatrix}
-
\begin{pmatrix}
U_N(x, x/\varepsilon, \varepsilon) \\ V_N(x, x/\varepsilon, \varepsilon)
\end{pmatrix}
=
-\varepsilon^N
(\mathfrak{M}(\varepsilon) - E )^{-1}
h(x, x/\varepsilon, \varepsilon)
\end{equation*}
where
\begin{equation}\label{EqWithResol}
h(x, x/\varepsilon, \varepsilon) =
\left[
(\mathfrak{M}_x - E )
\begin{pmatrix}
u_N(x, y) \\ v_N(x, y)
\end{pmatrix}
\right]_{y=x/\varepsilon}
+
(\mathfrak{M}(\varepsilon) - E ) \delta(x, \varepsilon).
\end{equation}

Using lemma~\ref{LmEstForAsimp}, (\ref{EstForDeU}), (\ref{EstForDeV}), and (\ref{EqWithResol}) we get
\begin{multline}\label{EstMainRes}
\left\|
(\mathfrak{M}(\varepsilon) - E )^{-1}
\begin{pmatrix}
f^u(x) \\ f^v(x)
\end{pmatrix}
-
\sum_{n=0}^{N-2} \varepsilon^n
\begin{pmatrix}
u_n(x, x/\varepsilon) \\ v_n(x, x/\varepsilon)
\end{pmatrix}
\right\|_{L_2(\mathbb{R}^3, \mathbb{C}^6)} \leqslant \\
\leqslant C_1 \varepsilon^{N-1}
\left\|
\begin{pmatrix}
f^u(x) \\ f^v(x)
\end{pmatrix}
\right\|_{H^{N+1}(\mathbb{R}^3, \mathbb{C}^6)},
\end{multline}
for $N \geqslant 2$, where the constant $C_1$ does not depend on $f^u(x)$ and $f^v(x)$.

Finally, we proved the following theorem.
\begin{theorem}\label{ThMain}
Suppose $\Im(E)>0$, $\varepsilon>0$, assumption {\rm \ref{assAlMu}} and conditions {\rm (\ref{AssFuv})} hold.
Then estimate {\rm (\ref{EstMainRes})} holds for $N \geqslant 2$, where the constant $C_1$ does not depend on $f^u(x)$ and $f^v(x)$.
\end{theorem}

\begin{corollary}\label{ClrMain}
Suppose $\Im(E)>0$, $\varepsilon>0$, assumption {\rm \ref{assAlMu}} and conditions {\rm (\ref{AssFuv})} hold.
Then
\begin{equation*}
\left\|
(\mathfrak{M}(\varepsilon) - E )^{-1} -
\Theta(\varepsilon) (\hat{\mathfrak{M}} - E )^{-1};
H^3(\mathbb{R}^3, \mathbb{C}^6) \rightarrow L_2(\mathbb{R}^3, \mathbb{C}^3)
\right\| \leqslant C_1 \varepsilon,
\end{equation*}
where constant $C_1$ does not depend on $\varepsilon$ and
$\Theta(\varepsilon)$ is a multiplication operator on the matrix-valued function
\begin{equation*}
\begin{pmatrix}
\alpha(x, x/\varepsilon) Z^u(x, x/\varepsilon) \Lambda^u(x)^{-1} & 0 \\
0 & \mu(x, x/\varepsilon) Z^v(x, x/\varepsilon) \Lambda^v(x)^{-1}
\end{pmatrix},
\end{equation*}
\begin{equation*}
Z^u = (\zeta_1, \zeta_2, \zeta_3), \quad
Z^v = (\xi_1, \xi_2, \xi_3).
\end{equation*}
\end{corollary}

\proof
For the proof it is sufficient to apply theorem \ref{ThMain} for $N=2$.~$\square$

\end{document}